\documentclass[aps,superscriptaddress,twocolumn,twoside,floatfix,pra,nofootinbib,a4paper]{revtex4-2}

\usepackage[final]{changes}

\usepackage{microtype}
\usepackage{times}
\usepackage{amsfonts}
\usepackage{amsmath}
\usepackage{amssymb}
\usepackage{amsthm}
\usepackage{multirow}
\usepackage[normalem]{ulem}
\usepackage[T1]{fontenc}
\usepackage{float}
\usepackage{mathtools}
\usepackage{bm}
\usepackage[UKenglish,cleanlook]{isodate}   
\usepackage{mathrsfs}
\usepackage{dsfont}

\usepackage{subfigure}

\usepackage{physics}

\usepackage{xcolor}
\definecolor{carmine}{RGB}{150,0,24}
\definecolor{mycolor1}{rgb}{1.00000,0.00000,1.00000}

\usepackage{natbib}
\usepackage[colorlinks=true,linkcolor=blue,citecolor=magenta,urlcolor=blue]{hyperref}

\newcommand{\bracket}[3]{\langle#1|#2|#3\rangle}

\usepackage{caption}
\usepackage{graphicx}
\usepackage{tikz-cd}

\begin{document}


\title{Prepare-and-measure scenarios with photon-number constraints}


\author{Carles Roch i Carceller}
\affiliation{Physics Department and NanoLund, Lund University, Box 118, 22100 Lund, Sweden}
\author{Jef Pauwels}
\affiliation{Department of Applied Physics, University of Geneva, 1211 Geneva, Switzerland.}
\affiliation{Constructor Institute of Technology (CIT), Geneva, Switzerland.}
\author{Stefano Pironio}
\affiliation{Laboratoire d'Information Quantique, CP 225, Universit\'e libre de Bruxelles (ULB),\\ Av. F. D. Roosevelt 50, 1050 Bruxelles, Belgium}
\author{Armin Tavakoli}
\affiliation{Physics Department and NanoLund, Lund University, Box 118, 22100 Lund, Sweden}

\begin{abstract}
	We study correlations in the prepare-and-measure scenario when quantum communication is constrained by \deleted{its} photon-number statistics. Such constraints are natural and practical control parameters for semi-device-independent certification in optical platforms. To analyse these scenarios, we \replaced{show how semidefinite programming relaxations for non-commutative polynomial optimization can be used}{introduce a semidefinite programming hierarchy} to bound the set of quantum correlations under restrictions on the photon-number \replaced{distribution}{components}. The practicality of this method is demonstrated by computing optimal performance bounds on several well-known communication tasks. We then \replaced{apply}{extend} the method \replaced{to the certification of semi-device-inpependent}{so that it applies to benchmarking} random number generation protocols \replaced{and show how to bound}{via bounds on} the conditional Shannon entropy. We showcase this versatile tool by improving randomness extraction in established protocols based on coherent states and homodyne measurements.
\end{abstract}


\maketitle


\paragraph*{Introduction.} The prepare-and-measure scenario is the standard setting for semi-device-independent (SDI) quantum information processing. The sender, Alice, encodes classical data into a quantum state relayed to Bob, who measures it based on a selected input. Using quantum systems in place of the classical systems unlocks a breadth of communication advantages \cite{Nayak1999, Gallego2010, Tavakoli2015, Brunner2013, Ahrens2012, Hendrych2012}. These advantages can be used indirectly to certify and benchmark quantum devices, and evaluate the performance of quantum information protocols in the SDI framework. This framework hosts a broad range of experiments with the common feature that they are uncharacterised up to only a limited assumption on the quantum states \cite{Pauwels2024}.  

A common SDI assumption is to limit the dimension of the quantum state. Dimension-restricted protocols have been developed for quantum key distribution \cite{Pawlowski2011, Woodhead2015}, random number generation \cite{Li2011, Li2012, Mironowicz2016}, self-testing \cite{Tavakoli2018, Farkas2019, Rosset2019, Mohan2019, Miklin2020, Miklin2021, IndepDevices} and entanglement detection \cite{FakeTriangle, Moreno2021, Bakhshinezhad2024}. They have also motivated a variety experiments \cite{Lunghi2015, Smania2020, Anwer2020, Foletto2020, Farkas2021, Martinez2018, Miao2022, Baumer2021}. However, the dimension assumption has important drawbacks: (i) multiple, incompatible, measurements are necessary \cite{Frenkel2015}, (ii) there is no dimension observable for verifying the assumption, and (iii) a fixed dimension is often only an approximation of a more complex system, which can undermine the validity of the model \cite{Pauwels2022}. 

Recently, several alternative SDI frameworks have emerged. These frameworks propose limiting quantum states based on various properties, such as their overlaps \cite{Brask2017, Wang2019}, information content \cite{Tavakoli2020informationally, Tavakoli2022informationally}, non-contextuality \cite{Flatt2022, Carceller2022}, control over state preparation \cite{Tav2021}, space-time symmetries \cite{Jones2023, Aloy2023}, or the expectation value of a physical observable \added{, such as the photon number operator or the projector on the vacuum component}\cite{VanHimbeeck2017semidevice}. Among these, constraining \replaced{photon-number statistics}{the states by a physical observable} is particularly practical—it avoids many of the challenges of other approaches, it does not require single-photon sources, and it is simple and cheap to verify. It has therefore attracted significant attention in both theoretical \cite{VanHimbeeck2017semidevice, vanhimbeeck2019correlations, senno2021semideviceindependent} and experimental studies \cite{Tebyanian2021, Rusca2020, Rusca2019b, Avesani2021}. However, existing protocols are restricted to scenarios where \replaced{constraints are placed only on a single parameter of the photon-number distribution, either the mean photon number or the $n$-photon component.}{the observable is chosen to be the projector onto the vacuum state,}\replaced{ Furthermore,}{and} only the simplest of such scenarios are currently solvable \cite{VanHimbeeck2017semidevice, vanhimbeeck2019correlations}.

Here, we advance this approach to SDI on three fronts, namely in terms of concepts, technical tools and protocol applications. Specifically, we go beyond the \replaced{previous assumptions}{vacuum component} and allow an experimenter to control more information about the photon-number statistics, i.e.~to \replaced{limit}{estimate} several \added{$n$-photon }components, or alternatively the truncated average photon number. This is particularly natural in the context of coherent states. Furthermore, we \replaced{show how}{develop} semidefinite programming (SDP) hierarchy techniques \cite{NPA2007,NPA2008,burgdorf2013tracial,burgdorf2016optimization,SDPreview} \added{can be used }for characterising the set of  correlations under arbitrary linear constraints on the photon statistics. Via several case studies, we show that it is a useful practical tool, performing optimally already at low relaxation level for well-known tasks,  for which no previous solution was known. Moreover, we \replaced{combine}{further develop} the SDP method \added{with the entropy relaxation technique of \cite{brown2023}} so that it can be used for quantum random number generation. We apply it first to show that knowledge of additional photon-number components leads to more randomness. Then, we revisit the randomness generation protocol proposed in \cite{VanHimbeeck2017semidevice, vanhimbeeck2019correlations} and tested in \cite{Tebyanian2021, Rusca2020, Rusca2019b, Avesani2021} where the analysis was limited to binary-outcome homodyne detection. Our methods allow the analysis of randomness based on Shannon entropy using multi-outcome binning of the quadratures, and we show that this leads to significantly better rates.

\paragraph*{Quantum communication restricted by photon statistics.} Consider a scenario where Alice and Bob are connected by a one-way optical channel. Alice selects an input $x$ and encodes it into the Fock space of \replaced{a single}{ one or more} quantum optical mode\deleted{s}, described by the quantum state $\rho_x$.
Alice then sends $\rho_x$ to Bob, who selects an input $y$ and performs a corresponding quantum measurement $\{M_{b|y}\}$ with outcome $b$. The parties can coordinate their actions through a shared classical random variable $\lambda$ with distribution $q_\lambda$. The correlations in the experiment are described by the conditional probability distribution $p(b|x,y)=\sum_\lambda q_\lambda \Tr\left(\rho_x^{(\lambda)} M_{b|y}^{(\lambda)}\right)$.

\deleted{We consider a bound...}
\added{The probability of finding the state $\rho_x^{(\lambda)}$ prepared by Alice in the $n$'th excited level is
\begin{equation}\label{energycon_ind}
	\bracket{n}{\rho_x^{(\lambda)}}{n}=1-\omega^\lambda_{x,n} \,.
\end{equation} 
We make the assumption that the \emph{average} probability of finding the $x$'th state in the $n$'th excited level is bounded, i.e.~, that
\begin{equation}\label{energycon-average}
	\sum_\lambda q_\lambda \omega^\lambda_{x,n} \leq \omega_{x,n} \,,
\end{equation}
for some $\omega_{x,n} \geq0$. Hence, $\omega_{x,n}$ represents a bound on the support of the average state $\rho_x := \sum_\lambda \rho_x^{(\lambda)} q_\lambda$ on the component with $n$ excitations:
\begin{equation}\label{energycon}
	\bracket{n}{\rho_x}{n}\geq 1-\omega_{x,n} \,.
   \end{equation}}
 Note that one could also consider the case where a restriction is placed on each individual state $\rho_x^{(\lambda)}$ \cite{VanHimbeeck2017semidevice}, i.e.~, at the level of (\ref{energycon_ind}), but this would be a stronger restriction, which \replaced{would require having access or knowledge to the states $\rho_x^{(\lambda)}$, potentially hidden to the users}{ cannot be verified via an observable}. We can select how many excited levels we wish to consider by putting a truncation\added{, i.e.~, } \replaced{$n\leq n_\text{trunc}$}{$n=n_\text{trunc}$}. In the simplest case of two preparations and a binary measurement, the state space corresponds to a qubit \cite{VanHimbeeck2017semidevice} and therefore \replaced{constraining a single component, such as the vacuum component, is already sufficient}{the only relevant constraint is the vacuum ($n_\text{trunc}=0$)} \cite{footnote1}. However, such simple reductions of the state and constraints are \replaced{exceptional cases}{pathological}.   

\replaced{The assumption on the $n$-photon components can stem from a direct experimental measurement at the output of the source, prior knowledge or characterization of the source, or a combination of both. For instance, using photon-number-resolving detectors—though experimentally demanding—one can directly estimate each individual \(\omega_{x,n}\). Alternatively, if the source is assumed to follow a Poisson distribution, the average photon number $\langle N_x \rangle$ can be experimentally measured, enabling the derivation of the individual \(n\)-photon components from the distribution.}{Estimating each individual $\omega_{x,n}$ in photonics means having available photon-number resolving detectors, which can be experimentally demanding.  The distribution can be assumed to follow Poisson statistics, fully characterized by the average photon number $\langle N_x \rangle$, which can then be estimated.} \replaced{Note that}{Alternatively, one can restrict to a quantity that is directly accessible experimentally, namely} the truncated number operator $\sum_{n=0}^{n_{\rm trunc}} n \bracket{n}{\rho_x}{n} \leq \langle N^\text{trunc}_x \rangle$\added{ is often directly accessible experimentally}. This is meaningful when combined with a limitation $\sum_{n=0}^{n_{\rm trunc}}  \bracket{n}{\rho_x}{n} \geq 1-\varepsilon_x$ \cite{Pauwels2022}. Here, $\varepsilon_x$ quantifies how much of the state is leaked to higher-order photon components. This is particularly simple for coherent states, which obey Poisson statistics, $P_x(n)=e^{-|\alpha_x|^2}\frac{|\alpha_x|^{2n}}{n!}$, where $\alpha_x\in\mathbb{C}$. Then one has $\langle N^\text{trunc}_x \rangle = \sum_{n=0}^{n_{\rm trunc}} n P_x(n)$ and $\varepsilon_x = 1-\sum_{n=0}^{n_{\rm trunc}} P_x(n)$.
We discuss this in more detail in Appendix~\ref{app:average}, and study an explicit example.

\paragraph*{Quantum correlations.} Consider that we are given a set of photon-number constraints $\{\omega_{x,n}\}$ for $n=0,\ldots,n_{\rm trunc}$, and are asked to select optimal states and measurements in order to maximise a generic linear functional\added{ on the probabilities}, $\mathcal{W} := \sum_{b,x,y}c_{bxy}p(b|x,y)$, for some real coefficients $c_{bxy}$. When \added{the underlying space is convex, which is the case when }assuming free shared randomness, such functionals characterise the set of correlations.  A simple heuristic for finding a lower bound on the optimal quantum value is to employ semidefinite programs (SDPs) in a seesaw \cite{SDPreview}. That is, we first select a random set of states and evaluate the optimal value of $\mathcal{W}$ as an SDP over the measurements of Bob. Fixing this solution, we compute the optimal states compatible with \eqref{energycon} as an SDP. This procedure is iterated until desirable convergence\footnote{It is necessary in this procedure to fix the Hilbert space dimension; an advisable choice is to set it equal to the number of state preparations \cite{Pauwels2024}.}.

The more interesting and less straightforward question is to determine upper bounds on $\mathcal{W}$, or more generally determine whether a given distribution $p(b|x,y)$ admits a quantum model compatible with given constraints on the photon statistics. We address this by
\replaced{rephrasing this problem as a non-commutative polynomial optimization (NPO) problem, which can be relaxed through the (unnormalized)  tracial variant \cite{burgdorf2013tracial,burgdorf2016optimization,gribling2019lower} of the associated hierarchy of SDP relaxations \cite{NPA2007,NPA2008,pironio2010convergent}}{developing a tracial variant of a hierarchy of SDP relaxations for the quantum set }.

\added{Let us first consider the case without shared randomness where the quantum model is defined by a set of states $\{\rho_x\}$ and measurements $\{M_{b|y}\}$, yielding $p(b|x,y)=\Tr\left(\rho_x M_{b|y}\right)$ and satisfying the constraints (\ref{energycon}).}
Let us define the set of $n_\text{trunc}+1$ operators $\sigma_n:=\ket{n}\bra{n}$ which correspond to projectors onto distinct photon-number states,\ i.e.~they satisfy $\Tr (\sigma_n)=1$ and  $\sigma_{n}\sigma_{n'}=\delta_{n,n'}\sigma_n$. Define a list of operators $O=\{\openone,\{\rho_x\}_x,\{M_{b|y}\}_{b,y},\left\{\sigma_n\right\}_n\}$ and let $S_k$ be the set of monomials over $O$ containing all products of length $k$. To this, we associate a $|S_k|\times |S_k|$ moment matrix, $\Gamma_{u,v}:=\Tr\left(u^\dagger v\right)$, for $u,v\in S_k$. By construction, we have $\Gamma\succeq 0$. 

We now make the following observations. (i) The probabilities appear in the moment matrix as $p(b|x,y)=\Gamma_{\rho_x,M_{b|y}}$. (ii) The $n$-photon component appears in the moment matrix as $\Gamma_{\rho_x,\sigma_n}$. Hence, the restrictions \eqref{energycon}  become linear constraints $\Gamma_{\rho_x,\sigma_n}\geq 1-\omega_{x,n}$. (iii) Several constraints on $\Gamma$ follow from normalisation of states and measurements, the properties of $\{\sigma_n\}$, the cyclicity of the trace and $M_{b|y}M_{b'|y}=\delta_{bb'}M_{b|y}$\footnote{The projectivity of the measurements can be assumed w.l.g.\replaced{, as Bob can always}{~because we can always permit Bob to} perform a Neumark dilation of his measurement.}. In general, Alice's states may be mixed even when conditioned on $\lambda$ because purifying them could increase the weight of their photon-number components. \replaced{The corresponding operators $\rho_x$ thus satisfy the semidefinite constraints $\rho_x-\rho_x^2 \succeq 0$ (instead of the more restrictive one $\rho_x-\rho_x^2 =0$ in the case of pure states). Such constraints can be accounted for by introducing localising matrices $\Upsilon_x$ with entries $\left(\Upsilon_x\right)_{u,v}:=\Tr\left(u^\dagger\left(\rho_x-\rho_x^2\right) v\right)$, for $u,v\in S_{k'}$, where the relaxation level $k'$ can be chosen freely. These matrices satisfy by construction $\Upsilon_x\succeq 0$.}{To permit mixed states, we impose the semidefinite constraint $\rho_x-\rho_x^2 \succeq 0$ via the localising matrices  $\Upsilon_{u,v}:=\Tr\left(u^\dagger\left(\rho_x-\rho_x^2\right) v\right)$, for $u,v\in S_{k'}$, where the relaxation level $k'$ can be chosen freely. The constraint is then relaxed to $\Upsilon\succeq 0$.}

In summary, a necessary condition for the existence of a quantum model for $p(b|x,y)$ compatible with given photon component restrictions is that there exist\deleted{s} positive semidefinite matrices $\Gamma$ and \replaced{$\Upsilon_x$}{$\Upsilon$} with the above properties, which is decided by SDP. If instead of deciding feasibility we are interested in optimising $\mathcal{W}$, then we introduce the objective function as a linear combination over the moment matrix, namely $\mathcal{W}=\sum_{b,x,y}c_{bxy} \Gamma_{\rho_x,M_{b|y}}$. The SDP hierarchy then returns upper bounds on $\mathcal{W}$. 

\added{We conclude by noting that the above SDP relaxations are convex in $p(b|x,y)$ and $\omega_{x,n}$ and thus shared randomness is implicitly taken into account.  That is if $\Gamma^{(\lambda)}$, $\Upsilon_x^{(\lambda)}$  are valid moment and localizing matrices associated to the correlations $p^{(\lambda)}(b|x,y)$, and the photon-number bounds $\omega_{x,n}^{(\lambda)}$, then the moment and localizing matrices $\Gamma = \sum_\lambda q_\lambda \Gamma^{(\lambda)}$ and $\Upsilon_x = \sum_\lambda q_\lambda \Upsilon_x^{(\lambda)}$ are valid for the correlations $p(b|x,y)= \sum_\lambda q_\lambda p^{(\lambda)}(b|x,y)$ and the photon-number bounds $\omega_{x,n} \geq \sum_\lambda q_\lambda \omega^{(\lambda)}_{x,n}$.}

\begin{figure*}
\centering
\includegraphics[width=\textwidth]{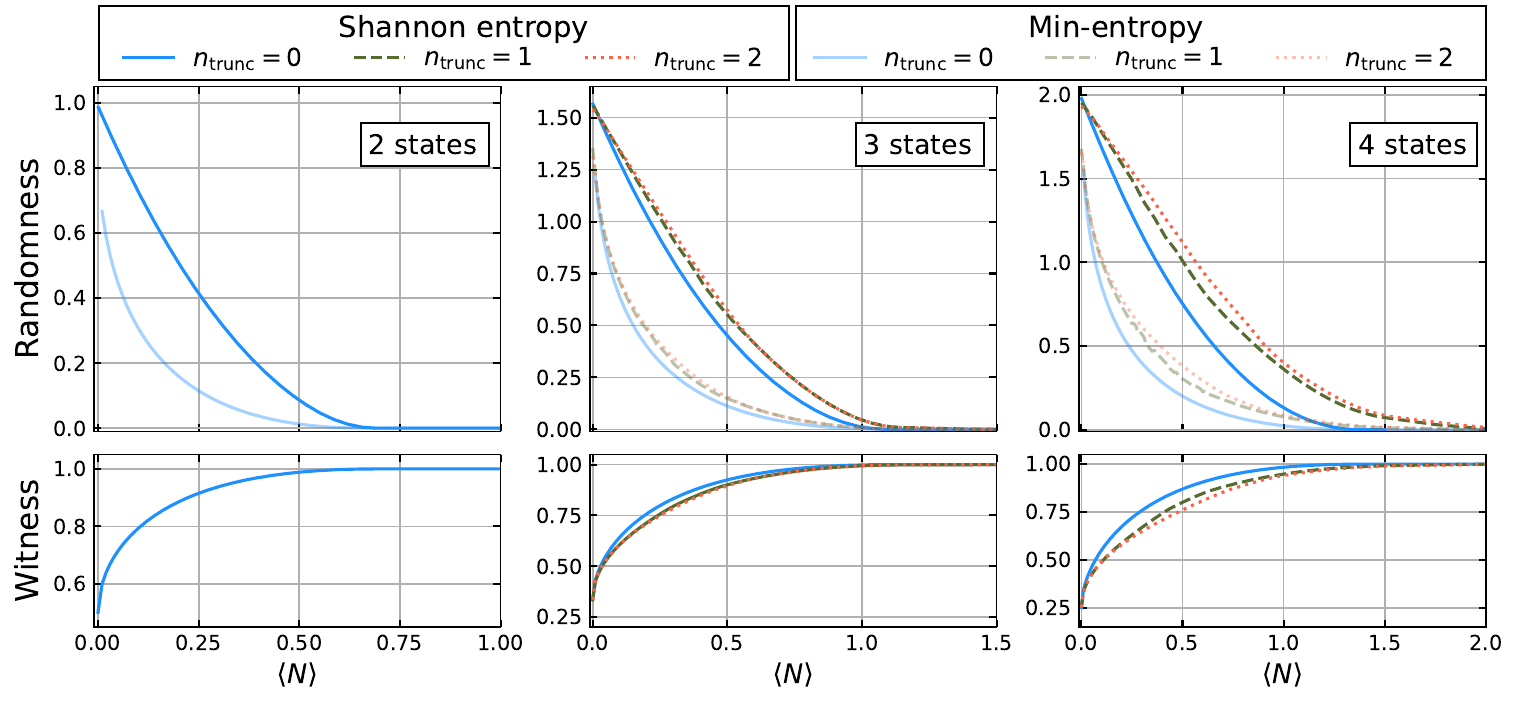}
    \caption{
			The success probability $\mathcal{W}_{n \rm disc}$ of discriminating between two, three and four quantum states \replaced{with the $n$-photon components with $n\leq n_{\rm trunc}=0,1,2$ (in blue, green, and red, respectively) fixed to match those of a Poisson distribution with mean photon number $\langle N \rangle$; no assumptions are made about the higher photon-number components.}{as a function of the average photon number $\langle N \rangle$ assuming Poisson statistics. We constrain $n_{\rm trunc}=0,1,2$ photons in the coherent state.} We also show the maximal randomness certified through the min- and Shannon entropy \added{at the maximum quantum value of the witness $\mathcal{W}_{n \rm disc}$}. \label{fig:disc}
}
\end{figure*}

\paragraph*{Application to communication tasks.} The central question is whether the  SDP relaxation method is useful in practice, and in particular whether it can give optimal bounds for relevant problems at reasonable computational cost. We will show an affirmative answer by considering well-known communication tasks in our framework.

We begin considering a simple two-state discrimination scenario. While Alice selects $x\in\{1,2\}$, Bob has a fixed measurement setting ($y=1$) with a binary outcome $b\in\{1,2\}$. The goal is to maximise the probability that Bob correctly discriminates Alice's state preparation, i.e.~$\mathcal{W}_\text{2 disc}=\frac{1}{2}p(1|1,1)+\frac{1}{2}p(2|2,1)$. For simplicity, we \replaced{impose the same bound on the vacuum component of the two states prepared by Alice and set}{name} $\omega_{x,0}=\omega$.  In \cite{VanHimbeeck2017semidevice} it was analytically proven that the optimal quantum protocol achieves $\mathcal{W}_{2~\text{disc}}^Q=\frac{1}{2}+\sqrt{\omega(1-\omega)}$. Using the SDP relaxation method, with a moment matrix corresponding to monomials $\{\openone, \rho M, \rho^2, \sigma_0 \rho,\sigma_0 M\}$, and a localizing matrix corresponding to monomials $\{\openone, \rho,M\}$, we recover the analytical result with seven decimal accuracy.
Then, we can go to scenarios where complete results are not known. Consider a general $n_X$-state discrimination situation, corresponding to success metric $\mathcal{W}_{n_X~\text{disc}}=\frac{1}{n_X}\sum_{x=1}^{n_X}p(x|x,1)$. We focus on the cases $n_X=3$ and $n_X=4$\added{, and again constrain only the vacuum component and identically for the different values of $x$, i.e., $\omega_{x,0}=\omega$}.
To the previous monomials we  add $\{\rho M \sigma_0,M\rho \sigma_0\}$. The resulting upper bounds are, in both cases, identical (up to $\sim 10^{-7}$), with lower bounds that we obtain from interior point search via the seesaw method. They correspond to $\mathcal{W}^Q_\text{3 disc}= \frac{1+\omega}{3}+\frac{2\sqrt{2}}{3}\sqrt{\omega(1-\omega)}$ and $\mathcal{W}^Q_\text{4 disc}= \frac{1+2\omega}{4}+\frac{\sqrt{3}}{2}\sqrt{\omega(1-\omega)}$. These bounds are identical to those derived analytically in \cite{Pauwels2024} but now without assuming pure states.

Recall that for the two-state discrimination protocol, the photon statistics are fully characterised by the vacuum component \cite{VanHimbeeck2017semidevice}. For three- and four-state discrimination, this is no longer true. To showcase the relevance of additional photon numbers, we constrain \replaced{the $n$-photon components for $n\leq n_\text{trunc}=0,1,2$}{photon-number components}, assuming the individual weights follow a Poisson distribution \replaced{characterized}{(fixed} by the average photon numbers $\added{\langle N_x \rangle=}\langle N \rangle$\deleted{)}\replaced{. We}{and} use SDP relaxations (at the same level) to bound the success rate as a function of $\langle N \rangle$.  The results are illustrated in Fig.~\ref{fig:disc} (bottom panel). They were matched up to at least four decimals via seesaw search.

Next, let us consider a scenario where Bob has inputs. We focus on the well-known quantum random access code, where Alice selects $x_0x_1\in\{00,01,10,11\}$, Bob chooses $y\in\{0,1\}$ with outcome $b\in\{0,1\}$. The goal is to output $b=x_y$. The average probability of success is $\mathcal{W}_{\text{RAC}}=\frac{1}{8}\sum_{x_0,x_1,y} p(b=x_y|x_0x_1,y)$. \replaced{We look for}{To find} the optimal quantum protocol when all four states have a \deleted{non-}vacuum weight \deleted{of} $\added{1-}\omega$\replaced{. One}{, one} must typically use different relaxation levels depending on the regime of $\omega$\footnote{E.g.~in the interval $0<\omega \leq \frac{1}{10}$, we extend our previous monomial list with the entries corresponding to $\{\sigma_0\rho\sigma_0,M^2\sigma_0,\rho M^2\}$.}. Our upper bounds are matched, up to in the worst case a precision $10^{-6}$, with lower bounds obtained from seesaw. In the interval  $0<\omega \leq \frac{1}{10}$, the upper bound can be approximated by $\mathcal{W}^Q_\text{RAC}\approx \frac{1}{2}+\sqrt{\frac{\omega}{2}}$ up to order $10^{-3}$.

\paragraph*{Random number generation.} 
A natural application of our framework is to SDI quantum random number generation (QRNG). We aim to certify randomness of Bob's outcomes \added{based only on the assumed photon-number restrictions and} with respect to any potential adversary that \replaced{has classical-side information about}{is classically correlated with} the devices \deleted{based only on the assumed photon-number restrictions}. \added{Classical side-information means that the adversary is not entangled with the quantum systems of the devices but is only classically correlated with them, e.g.~, through pre-shared classical variables $\lambda$. This assumption is satisfied if the devices are assumed to have no quantum memory, a reasonable and realistic assumption in the semi-device-independent setting and given the current state of technology.}
\deleted{As in the previous, Alice's and Bob's devices can at most pre-share a classical random variable....}

\added{Denote by \(\mathcal{Q}(\{\omega_{x,n}\})\) the set of quantum correlations \(p(b|x,y)\) compatible with the photon-number constraints (\ref{energycon}) defined by the bounds \(\{\omega_{x,n}\}\). Assume that Alice and Bob’s devices are characterized by a given probability distribution \(p = \{p(b|x,y)\}\). In our semi-DI setting, the randomness of Bob's output when using the inputs \(X = x^*\) for Alice and \(Y = y^*\) for Bob can be quantified through the following conditional Shannon entropy:}
\begin{align}\label{eq:shannon}
	H(B|X = &x^*, Y = y^*, \Lambda)  = \nonumber \\
	\min_{\{q_\lambda, p^{(\lambda)}\}} &\quad  -\sum_\lambda q_\lambda \sum_b p(b|x^*, y^*, \lambda) \log p(b|x^*, y^*, \lambda) \,, \nonumber \\
	\text{s.t.} & \quad  \sum_\lambda q_\lambda p^{(\lambda)} = p \,, \\ 
	& \quad \sum_\lambda q_\lambda \omega^{(\lambda)}_{x,n} \leq \omega_{x,n} \,, \nonumber \\
	 & \quad  p^{(\lambda)} \in \mathcal{Q}(\{\omega^{(\lambda)}_{x,n}\}) \,, \nonumber \\ 
	 & \quad  q_\lambda \geq 0 \,, \quad \sum_\lambda q_\lambda = 1 \,, \nonumber
\end{align}
\added{where the minimum is taken over the possible ensembles \(\{q_\lambda, p^{(\lambda)}\}\) of decompositions of the observed correlations $p$ in term of the hidden variables $\lambda$ that are compatible with the given average energy bounds $\omega_{x,n}$. This optimization provides a lower bound on the randomness of Bob's outcomes, measured by the conditional Shannon entropy, under the assumption that the adversary has access only to classical side-information \(\Lambda\).}

\added{
The above entropy is a single-round measure that characterizes the probabilistic behavior of the devices in an individual run. In practice, a randomness generation protocol involves \(n\) such runs. However, the devices may not behave in an i.i.d. (independent and identically distributed) manner, and only frequencies of outcomes—not the underlying probabilities—are directly observable. As shown in \cite{vanhimbeeck2019correlations}, a general protocol that includes a statistical estimation step and does not rely on the i.i.d. assumption can extract approximately
$ n H(B|X = x^*, Y = y^*, \Lambda)$
bits of randomness (up to corrections of order \(\sqrt{n}\)) from devices whose ideal behavior is expected to follow the distribution \(p(b|x, y)\). The analysis in \cite{vanhimbeeck2019correlations} is inspired by previous work on device-independent randomness certification \cite{arnon2018practical,arnon2019simple,zhang2018certifying,knill2020generation}, but importantly also accounts for constraints on the average value of several physical observables, such as the \(n\)-photon projectors in our case.  For further details, we refer the reader to \cite{vanhimbeeck2019correlations}.}

\added{
Consequently, the task of randomness certification reduces to computing bounds on the above single-round Shannon entropy. The set of quantum correlations, as discussed previously, admits NPO formulation that can be relaxed to a hierarchy of SDPs. The primary challenge, however, arises from the objective function—the Shannon entropy—which is non-linear in the probabilities and therefore not directly amenable to standard SDP techniques.}

\added{
A straightforward approach to address this issue is to instead optimize the min-entropy, which provides a lower bound on the Shannon entropy. This can be achieved by replacing the objective function with:
\begin{equation}
	\begin{split}
	H_{\rm min}(B&|X = x^*, Y = y^*, \Lambda) = \\ 
	&-\log \min_{\{q_\lambda, p_\lambda\}} \sum_\lambda q_\lambda \max_b p(b|x^*, y^*, \lambda) \,.
	\end{split}
\end{equation}
Optimizing the min-entropy can be done as shown in \cite{Silleras2014, Bancal2014} by noting that one can restrict to a finite number of hidden variables \(\lambda\), with one hidden variable for each \(b\). Specifically, for \(\lambda = b\), we have \(\max_b p(b|x^*, y^*, \lambda = b) = p(b|x^*, y^*, \lambda = b)\). This simplification makes the objective function linear in the hidden probabilities \(p(b|x^*, y^*, \lambda = b)\), reducing the problem to an optimization over a finite number of parameters. Since both the constraints and the objective are now linear, the problem can be fully formulated and solved as a finite SDP, providing a computationally tractable way to certify randomness. }

\added{
While the min-entropy provides a lower bound on the Shannon entropy, better estimates of randomness can, in principle, be obtained by directly bounding the Shannon entropy itself. This can be achieved by combining our SDP relaxation of the quantum set with the Brown-Fawzi-Fawzi (BFF) method \cite{brown2023}. 
The BFF method relies on an integral representation of the logarithm, which is discretized using Gauss-Radau quadratures. In its general form, the method is designed to bound the von Neumann entropy and introduces additional operators \(Z_{i,b}\), where each quadrature point \(i\) and outcome \(b\) is associated with a separate operator in the NPO formulation. We specialize the method to the Shannon entropy, which amounts top assume that the \(Z_{i,b}\) operators are scalar (see \cite{carceller2024} for a similar approach). This modification adapts the BFF method to directly certify Shannon entropy while leveraging the structure of our SDP relaxation.  In particular, we take advantage of the convex structure of the problem, which allows us to efficiently handle the in-principle arbitrary number of hidden parameters $\lambda$.
The technical details of this approach are provided in Appendix~\ref{app:shannon}.}

\added{We, remark that in the SDP formulations the constraint \(\sum_\lambda q_\lambda p_\lambda = p\) can be relaxed to require only that the ensemble \(\{q_\lambda, p_\lambda\}\) reproduces specific statistical properties of the observed distribution $p$. For instance, instead of matching the full probability distribution, the ensemble could be constrained to satisfy conditions on the expected value of a witness or other moments of the distribution. Similarly, the bounds on the \(n\)-photon components can be replaced with more general linear constraints on these components. For example, one might impose bounds on the expectation value of the truncated number operator, i.e., $\sum_{n=0}^{n_{\rm trunc}} n \bracket{n}{\rho_x}{n} \leq \langle N^\text{trunc}_x\rangle$ or other relevant quantities.}

Finally, note that using this SDP method one can also extract a trade-off function from the dual program which allows for finite-statistics analysis.

\begin{figure}[t!]
	\centering
	\includegraphics[width=0.95\columnwidth]{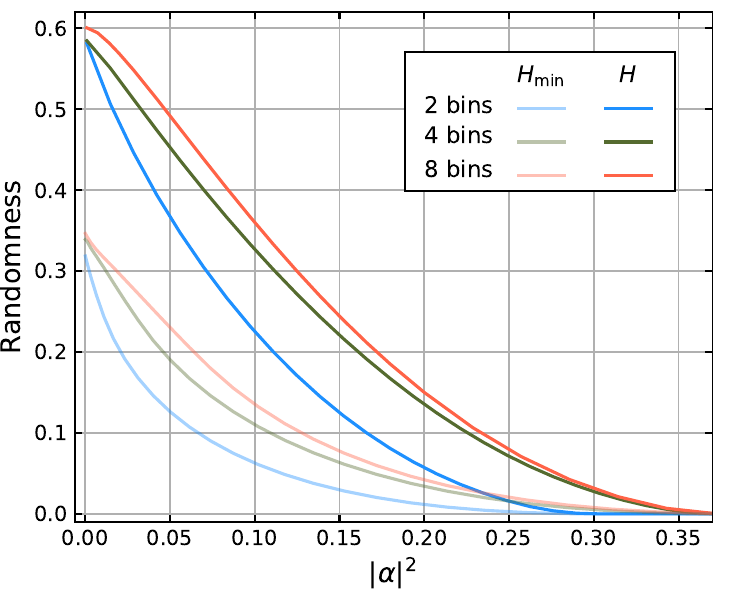}
	\caption{Randomness from the BPSK protocol as a function of the average photon number $\left|\alpha\right|^2$, using different binnings and bounding the vacuum component ($n_{\rm trunc}=0$). Significantly more randomness can be extracted from the distribution when binning to more outcomes. \label{fig:BPSK}}
\end{figure}

\paragraph*{QRNG protocols.} To showcase these methods, we apply them to two illustrative examples. First, we consider the state-discrimination witness introduced before (see Figure~\ref{fig:disc}) and investigate the effect of taking into account more photon number \added{constraints} ($n_{\rm trunc}=0,1,2$) on the certifiable randomness. As before, we assume Poisson photon statistics, characterised by the average photon number $\langle N \rangle$, and set the security parameter $\mathcal{W}_{n_\text{X} ~ \rm disc}$ to equal $\mathcal{W}_{n_\text{X} ~\rm disc}^{Q}$, where $\mathcal{W}_{n_\text{X} ~\rm disc}^{Q}$ is the maximal $\mathcal{W}_{n_\text{X} ~ \rm disc}$ compatible with the constraints. We then compute both the min- and Shannon entropy\footnote{For the min-entropy, we consider relaxations at the same level we as before. For the Shannon entropy, we consider moment matrix relaxations at level $k=2$, localizing matrices at level $k'=1$ and Gauss-Radau quadratures up to $m=8$}. The resulting rates are plotted in Fig.~\ref{fig:disc}. With seesaw methods \cite{carceller2024}, we verified that these rates are at worst nearly optimal. As expected, considering more photon number constraints leads to significantly higher rates both based on the min- and Shannon entropy. We obtain nearly the maximal randomness rate of $\log(n_X)$ bits in the limit of vanishing energies in all cases. The Shannon entropy gives a significant improvement over the min-entropy and it is less sensitive to small fluctuations in the photon component weights.

Next, we turn to a more practical example, and apply our methods to the Binary Phase-Shift Keying (BPSK) protocol, introduced in \cite{vanhimbeeck2019correlations,VanHimbeeck2017semidevice} and experimentally demonstrated in \cite{Rusca2020}; see also \cite{Rusca2019b, Tebyanian2021}. Alice prepares one of two coherent states $\ket{\pm \alpha}$, where $\alpha>0$ real. Both states have average photon number $\langle N \rangle = \alpha^2$, and non-vacuum component $1-\exp(\alpha^2)$. Bob performs a homodyne measurement of the $X$ quadrature. Refs.~\cite{vanhimbeeck2019correlations,VanHimbeeck2017semidevice} consider only a binary binning of the continuous outcome, $b = \text{sgn}(X)$ because this permits an solution based on a qubit state space. They leave as an open problem whether the rate can be improved by considering more fine-grained binning. Our methods readily enable an answer this question\footnote{We note that the improvement cannot be arbitrarily high since the number of inputs is finite \cite{Ioannou2019}.}. In addition to binary binnings, we consider four- and eight-outcome binnings. Our choice of binning intervals and the resulting probability distributions are given in Appendix~\ref{App:BPSK}. We then compute the min- and Shannon entropy based on the full distribution and a vacuum restriction ($n_{\rm trunc}=0$)\footnote{We consider moment matrix relaxations at level 
$k=2+\rho\rho\rho + \sigma M \rho$, localising matrices at level $k'=1$ and Gauss-Radau quadratures up to $m=8$.}. The results are illustrated in Fig.~\ref{fig:BPSK}. For binary binnings, comparison of our results to the Shannon entropy computed in \cite{vanhimbeeck2019correlations} using semi-analytic methods shows agreement up to numerical precision, confirming that our results are nearly optimal. Furthermore, we find that increasing the number of bins significantly boosts the randomness. For instance, at $\alpha^2=0.1$, the rate increases by approximately $\sim 150\%$ and $\sim 200\%$ for the four- and eight-outcome binnings, respectively.  Finally, we again see that the Shannon entropy provides a higher estimate of the randomness than does the min-entropy.

In the analysis above, we focussed on component-wise constraints of the type \eqref{energycon}. In Appendix~\ref{app:average}, we study how our results change when bounding instead a weighted average of the photon numbers, corresponding to a (truncated) photon number operator. We find that at low energies, the resulting rates are qualitatively similar, while at higher energies the rates are significantly reduced. This is expected, as the average photon number is a less stringent constraint than the individual photon number components.

\paragraph*{Conclusions.} We have introduced a theoretical framework and computational tools for quantum correlations under arbitrary photon number restrictions or weighted linear combinations of them. This opens up several possibilities for applications in semi-device-independent quantum information, both for enhancing the performance of previous experiments and for novel applications in quantum communication and cryptography. Interesting future directions are the application of these methods to develop novel continuous-variable quantum key distribution protocols, more practical randomness generation protocols and to understand the role of entanglement between Alice and Bob. 



\begin{acknowledgments}
C.R.C and A.T. are supported by the Wenner-Gren Foundation, by the Knut and Alice Wallenberg Foundation through the Wallenberg Center for Quantum Technology (WACQT) and the Swedish Research Council under Contract No.~2023-03498. J.P. is supported by NCCR-SwissMAP. S.P. acknowledges funding from the VERIqTAS project within the QuantERA II Programme that has received funding from the European Union's Horizon 2020 research and innovation program under Grant Agreement No 101017733 and the F.R.S-FNRS Pint-Multi program under Grant Agreement R.8014.21, from the European Union's Horizon Europe research and innovation program under the project ``Quantum Security Networks Partnership'' (QSNP, grant agreement No 101114043),  from the F.R.S-FNRS through the PDR T.0171.22, from the FWO and F.R.S.-FNRS under the Excellence of Science (EOS) program project 40007526, from the FWO through the BeQuNet SBO project S008323N, from the Belgian Federal Science Policy through the contract RT/22/BE-QCI and the EU ``BE-QCI'' program.
S.P. is a Research Director of the Fonds de la Recherche Scientifique - FNRS. 

Funded by the European Union. Views and opinions expressed are however those of the authors only and do not necessarily reflect those of the European Union. The European Union cannot be held responsible for them.

\paragraph*{Code availability.} The code used to generate the results in this paper is available on GitHub: \url{https://github.com/chalswater/Energy_restricted_QRNG}.
\end{acknowledgments}	

\bibliography{references_energy}

\appendix

\onecolumngrid

\section{Bounding the truncated photon-number operator} \label{app:average}

In the main text, we focussed on analyzing correlations and randomness based on constraints on individual photon components. Depending on the application, one may wish to relax this assumption, and consider instead a constraint on the truncated average photon-number operator. Our method can readily be adapted to accommodate this case (and in fact can be used to bound any weighted average of the photon-number operators). Here, we study an illustrative example where we bound the truncated average photon-number operator,
\begin{equation} 
\sum_{n=0}^{n_{\rm trunc}} n \bracket{n}{\rho_x}{n} \leq \langle N^\text{trunc}_x \rangle \,, 
\end{equation}
where $\langle N^\text{trunc}_x \rangle$ is the average photon number in the truncated space corresponding to $n \leq n_{\rm trunc}$.

The above constraint by itself does not impose any constraints on the achievable correlations. Indeed, all relevant quantum information may be encoded in higher photon numbers $n > n_{\rm trunc}$. We thus need to impose an additional constraint reflecting the assumption that most of the weight of the states is contained in  the relevant subspace $n \leq n_{\rm trunc}$.
\begin{equation} 
\sum_{n=0}^{n_{\rm trunc}}  \bracket{n}{\rho_x}{n} \geq 1-\varepsilon_x \,,
\end{equation} 
where $\varepsilon_x$ 
quantifies how much of the state leaks to higher photon numbers. For $\varepsilon_x=0$, this constraint amounts to a dimension bound, i.e. the state is fully contained in the truncated space. For $\varepsilon_x>0$, the state is allowed to leak to higher photon numbers, i.e. it is an almost $n_{\rm trunc}$-dimensional state in the sense of Ref.~\cite{Pauwels2022}. One may even generalise further and choose this dimension to be different from the truncation level $n_{\rm trunc}$, which would allow for a more flexible description of the state.

Consider an example in which the experimenter assumes the source to be approximately Poissonian. She does not want to assume the entire photon statistics, because an adversary might be able to manipulate the low end of the distribution, but trusts that the leakage to higher photon numbers is what she expects for a Poissonian distribution. In addition, she can measure the average photon number (she has a power meter), but does not have a photon-number resolving detector. In that case, she may assume that:
$\langle N^\text{trunc}_x \rangle = \sum_{n=0}^{n_{\rm trunc}} n P_x(n)$ and $\varepsilon_x = 1-\sum_{n=0}^{n_{\rm trunc}} P_x(n)$. Another less practical but theoretically appealing example is to assume a dimension bound in combination with an effective dimension constraint, corresponding to the case $\varepsilon_x = 0$. 

\begin{figure}
	\centering
	\includegraphics[width=0.5\textwidth]{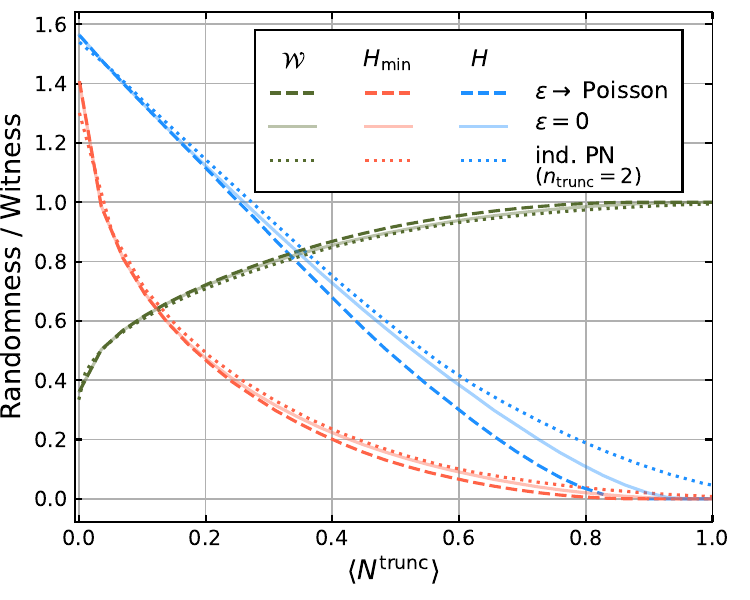}
	\caption{Randomness from the three-state discrimination game with bounded truncated average photon number for $n_{\text{trunc}}=2$. We consider the cases of zero leakage ($\varepsilon_x=0$) and the expected leakage from Poissonian statistics ($\varepsilon_x = 1-\sum_{n=0}^{n_{\rm trunc}} P_x(n)$). We additionally compare these results with the state discrimination games with individual photon-number (denoted as ``ind. PN'' in the figure) constraints (with $n_{\text{trunc}}=2$).}
	\label{fig:avE}
\end{figure}

We study both of these examples for the 3-state discrimination protocol and compute the Shannon and min-entropies in the outcomes based on the security parameters $\mathcal{W}_\text{3 disc}$. We compare to the case studied in the main text where we bound each of the photon-number components individually.
The results are shown in Fig.~\ref{fig:avE}. 

Restricting individual photon components is a stronger assumption than a bound on the average photon population, and indeed it translates to greater certifiable randomness. Nevertheless, this difference decreases as we reduce the energy of the source (i.e. when state preparations are less distinguishable), leading to almost identical entropy bounds for both assumptions. In this regime, the certifiable randomness is greater. Hence, one can relax the assumption to the truncated average photon number without paying a significant price in terms of reduced entropy.

\section{Bounding the Shannon entropy} \label{app:shannon}

Here, we present a derivation of the SDP relaxation for the Shannon entropy used in the main text.
We begin by considering the scenario where we aim to bound the Shannon entropy
\begin{equation}\label{eq:shannon_noE}
H(B|X=x^*,Y=y^*) = H(B|x^*,y^*) = -\sum_{b} p(b|x^*,y^*) \log p(b|x^*,y^*) \,,
\end{equation}
of the output distribution $p(b|x^*,y^*) = \Tr(\rho_{x^*}M_{b|y^*})$ for a specific choice of settings $x^*$ and $y^*$. In this case, we have implicitly excluded the eavesdropper, Eve, from consideration -- or equivalently, we assume that Eve has no side-information correlated with the devices used by Alice and Bob. The more general case (\ref{eq:shannon}), where Eve is classically correlated with Alice and Bob through shared randomness, will be addressed using a convexity argument to incorporate the influence of such correlations on the entropy bounds.

\textbf{Bound on $H(B|x^*,y^*)$ from \cite{brown2023} --}
Even though Eve is absent from the picture for now, it is useful to make a connection with the more general approach \cite{brown2023} for bounding the von Neumann entropy in the presence of arbitrary quantum side information. To do so, we can view our classical probability distribution as formally being associated with the classical-quantum (cq-) state:
\begin{equation}
    \sigma_{B Q_E} = \sum_b |b\rangle\langle b| \otimes p(b|x^*, y^*),
\end{equation}
where $ Q_E $ denotes Eve's register and is one-dimensional. Since the Shannon entropy $H(B|x^*, y^*)$ defined in (\ref{eq:shannon_noE}) is the conditional von Neumann entropy $H(B|Y=y^*,Q_E)$ of the above cq-state, this representation allows us to directly obtain bounds on the Shannon entropy \( H(B|x^*, y^*) \) as a special case of the von Neumann entropy bounds developed in \cite{brown2023} when applied to $\sigma_{BQ_E}$. This cq-state can be understood as the state obtained after Bob has performed his measurement on the initial quantum state:
\begin{equation}
    \rho_{Q_B Q_E} = \rho_{x^*} \otimes 1,
\end{equation}
where $ \rho_{x^*}$ represents the quantum state sent by Alice's and $1$ denotes the trivial one-dimensional state associated with Eve's register. Lemma~2.3 and eqs~(23) of \cite{brown2023} applied to this special case immediately yield the follewing bound on the Shannon entropy
\begin{equation}
    H(B|x^*, y^*) \geq c_m + \sum_{i=1}^{m-1}\tau_i \sum_{b} \ \underset{z^{b,i}}{\inf} \left\{ \Tr\left[\rho_{x^{*}} M_{b|y^\ast}\right]\left(z_{b,i}+ \bar{z}_{b,i}+(1-t_i)(\bar{z}_{b,i}z_{b,i})\right)+ t_i (\bar{z}_{b,i}z_{b,i}) \right\}\,,
\end{equation}
where $\tau_{i}:=\frac{w_{i}}{t_{i}\ln{2}}$ and $c_m:=\sum_{i=0}^{m-1}\tau_i$ are defined in term of the nodes $t_i$ and weights $w_i$of the Gauss-Radau quadratures (see \cite{Gautschi2000,Brown2021}), and $z_{b,i}$ are complex scalar variables. 
Since the right-hand side of the above expression is minimized when the $z_{b,i}$ are real, we can restrict them to be real without loss of generality. This allows us to rewrite the bound as
\begin{equation}\label{eq:sb}
    H(B|x^*, y^*) \geq c_m + \sum_{i=1}^{m-1}\tau_i \sum_{b} \ \underset{z_{b,i}}{\inf} \left\{ \Tr\left[\rho_{x^{*}} M_{b|y^\ast}\right]\left(2z_{b,i}+ (1-t_i)z_{b,i}^2\right)+ t_i z_{b,i}^2 \right\}\,.
\end{equation}

\textbf{Alternative derivation --}
An alternative derivation of the same bounds on the Shannon entropy can be obtained as follows. First, we express the base 2 logarithm $\log(u) = \ln(u)/\ln(2)$ in (\ref{eq:shannon_noE}) in term of the natural logarithm $\ln(u)$ and use the integral representation
\begin{equation}
    \ln(u) = \int_{0}^{1} \frac{u-1}{t(u-1)+1}dt\,,
\end{equation}
to write the Shannon entropy as 
\begin{equation}
    H(B|x^*, y^*) = \sum_{b} p(b|x^*,y^*) \frac{1}{\ln 2} \int_{0}^{1} \frac{1-p(b|x^*,y^*)}{t(1-p(b|x^*,y^*)) + p(b|x^*,y^*)}dt\,.
\end{equation}
Then, we discretize the integral using the Gauss-Radau quadratures to obtain the bound
\begin{equation}
H(B|x^\ast, y^\ast) \geq \sum_{b} p(b|x^{*}, y^{*})  \sum_{i=1}^{m-1} \frac{w_i}{\ln 2} \frac{1-p(b|x^{*}, y^{*})}{t_i(1-p(b|x^{*}, y^{*})) + p(b|x^{*}, y^{*})},.
\end{equation}
After some algebra, one obtains:
\begin{equation}
H(B|x^\ast, y^\ast) \geq \sum_{b} p(b|x^{*}, y^{*}) \sum_{i=1}^{m-1} \frac{w_i}{t_i \ln 2} \left( 1 - \frac{p(b|x^{*}, y^{*})^2}{t_i(1-p(b|x^{*}, y^{*})) + p(b|x^{*}, y^{*})} \right),
\end{equation}
which can be rewritten as:
\begin{equation}\label{eq:shannon_bound_fbi}
H(B|x^\ast, y^\ast) \geq c_m + \sum_{i, b} \tau_i \left( 1 - \frac{p(b|x^{*}, y^{*})^2}{t_i(1-p(b|x^{*}, y^{*})) + p(b|x^{*}, y^{*})} \right) = c_m + \sum_{i, b} \tau_i f_{b,i},
\end{equation}
where we defined
\begin{equation}
f_{b,i}:= -\frac{p(b|x^{*}, y^{*})^2}{t_i(1-p(b|x^{*}, y^{*})) + p(b|x^{*}, y^{*})}.
\end{equation}
The above expressions $f_{b,i}$ are of the rational form $f = -\frac{p^2}{t(1-p)+p}$. This type of expressions often arises as the minimum of a quadratic function $f = \min_z g(z) = \min_z \left(az^2 + bz\right)$. Indeed, the minimum of $g(z)$ occurs at $z^* = - \frac{b}{2a}$ if $a>0$. The minimum value is then $g(z^*) = -\frac{b^2}{4a}$. Thus, taking $a = p(1-t)+t$ and $b=2p$, we can write $f_{b,i}$ as the minimum of the following quadratic function:
\begin{equation}
f_{b,i} = \min_{z_{b,i}} p(b|x^{*}, y^{*}) \left[2z_{b,i} + (1-t_i) z_{b,i}^2 \right] + t_i z_{b,i}^2\,.
\end{equation}
Inserting this expression in the bound (\ref{eq:shannon_bound_fbi}), and using $p(b|x^*, y^*) = \Tr\left[\rho_{x^*} M_{b|y^*}\right]$, we find the same bound on the Shannon entropy (\ref{eq:sb}) already obtained through a direct application of \cite{brown2023}.

\textbf{SDP relaxation --}
Now given a set of linear constraints on the correlations $p(b|x,y)$ of the form $\sum_{bxy}\alpha^i_{bxy}p(b|x,y)\leq \alpha^i_0$, the search for a lower-bound on the Shannon entropy valid for any quantum realization compatible with these linear constraints and with given photon-number constraints can be formulated, using (\ref{eq:sb}), as the following NPO:
\begin{align}
    \label{eq:vn_cc}
       H(B|x^*,y^*)\geq \underset{\left\{\rho_x, M_{b|y},\sigma_n,Z_{b,i}\right\}}{\text{minimize}} & \quad c_m + \sum_{i=1}^{m-1} \tau_i \sum_{b}  \Tr\left[\rho_{x^{*}} \left(M_{b|y^*} (2 Z_{b,i} + (1-t_i)Z_{b,i}^2)\right) + t_i Z_{b,i}^2\right]  \\
       & \quad \Tr\left[\rho_x\right] = 1, \quad \rho_x-\rho_x^2 \succeq 0, \nonumber \\
       & \quad \sum_{b} M_{b|y} = \mathds{1}, \quad M_{b|y}M_{b'|y}=\delta_{bb'}M_{b|y}, \nonumber\\
       & \quad \Tr\left[\sigma_n\right] = 1, \quad \sigma_n\sigma_{n'} = \delta_{nn'}\sigma_n, \nonumber\\
       &\quad [Z_{b,i},Z_{b',i'}]=[Z_{b,i},\rho_x] =[Z_{b,i},M_{b'|y}]=[Z_{b,i},\sigma_n]=0,\nonumber\\
       &\quad \Tr\left[\rho_x \sigma_n\right] \geq 1 - \omega_{x,n},\nonumber \\
       &\quad \sum_{bxy}\alpha^i_{bxy} \Tr\left[\rho_x M_{b|y}\right] \leq \alpha^i_0 \nonumber\,
\end{align}
in term of the operator variables $\rho_x, M_{b|y}, \sigma_n$ and $Z_{b,i}$ where we interpret the scalar variables $z_{b,i}$ in (\ref{eq:sb}) as operator $Z_{b,i} = z_{b,i}\mathds{1}$ proportional to the identity, hence commuting with all other operators.

The above NPO problem can be immediately relaxed to a hierarchy of SDP's using the standard tracial moment matrix approach \cite{burgdorf2013tracial,burgdorf2016optimization} in its non-normalized version \cite{gribling2019lower}. That is, as explained in the main text, we start by defining the set $S_k$ of monomials of length $k$ in the operators $O=\{\openone,\{\rho_x\}_x,\{M_{b|y}\}_{b,y},\left\{\sigma_n\right\}_n,\left\{Z_{b,i}\right\}_{b,i}\}$. 
We then associate a $|S_k|\times |S_k|$ moment matrix $\Gamma$ with entries $\Gamma_{u,v}=\Tr\left(u^\dagger v\right)$, for $u,v\in S_k$ and $|S_{k'}|\times |S_{k'}|$ localizing matrices ${\Upsilon}_x$, with entries $\left({\Upsilon}_x\right)_{u,v}:=\Tr\left(u^\dagger (\rho_x-\rho_x^2)v\right)$, for $u,v\in S_{k'}$. By construction, we have $\Gamma\succeq 0$ and ${\Upsilon}_x \succeq 0$. The definition of these moment matrices, the cyclicity of the trace, and the operator and tracial equalities and inequalities appearing in (\ref{eq:vn_cc}) translate into linear constraints on the entries of $\Gamma$ and ${\Upsilon}_x$ in the standard way. 
The objective function in (\ref{eq:vn_cc}) can also be written as a linear function of the entries of $\Gamma$, specifically as 
\begin{equation}
    f(\Gamma) = c_m + \sum_{i=1}^{m-1} \tau_i \sum_{b} \left[2\Gamma_{Z_{b,i}\rho_{x^\ast},M_{b|y^\ast}} + (1-t_i)\Gamma_{Z_{b,i}\rho_{x^\ast},Z_{b,i}M_{b|y^\ast}} + t_i \Gamma_{Z_{b,i}\rho_{x^\ast},Z_{b,i}}\right]
\end{equation}
and the optimization problem can thus be relaxed as an SDP.

\added{
An useful constraint that we can add to the standard tracial SDP relaxations is that
\begin{equation}\label{eq:identitytrace}
    \Tr[\mathbf{Z} \rho_x] = \Tr[\mathbf{Z} \rho_{x'}] = \Tr[\mathbf{Z} \sigma_n] = \Tr[\mathbf{Z} \sigma_{n'}]\quad \forall x,x',n,n',
\end{equation}
where $\mathbf{Z}$ is an arbitrary product of the operators $Z_{b,i}$. The above constraint would not hold in general if the $Z_{b,i}$ were merely operators commuting with all the other operators. But here, we know that they satisfy the stronger condition $Z_{b,i} = z_{b,i}\mathds{1}$, which implies (\ref{eq:identitytrace}).}

\textbf{Lower-bounds on the Shannon entropy with classical side information --}
Consider now the case where Alice, Bob, and Eve are correlated through shared randomness $\lambda$, i.e., Eve has classical side-information about Bob's outcomes. For each $\lambda$, arising with probability $q(\lambda)$, Alice and Bob's use some quantum strategy, associated with moment and localizing matrices $\Gamma^{(\lambda)}$ and ${\Upsilon}^{(\lambda)}_x$. These moment and localizing matrices satisfy all the constraints arising from the NPO (\ref{eq:vn_cc}) \added{and (\ref{eq:identitytrace})} but with the weaker conditions $\omega_{x,n} \rightarrow \omega_{x,n}^{(\lambda)}$ and $\alpha_0^{i} \rightarrow \alpha_0^{i,(\lambda)}$ where $\sum_\lambda q(\lambda) \omega_{x,n}^{(\lambda)} \leq \omega_{x,n}$ and $\sum_\lambda q(\lambda) \alpha_0^{i,(\lambda)} = \alpha_0^{i}$, since the 
photon-number bounds $\omega_{x,n}$ and the linear constraints on the correlations should only hold on average. 
The conditional Shannon entropy is now $H(B|x^*,y^*,\Lambda) = \sum_\lambda q(\lambda) H(B|x^*,y^*,\lambda) = \sum_\lambda q(\lambda) \left(- \sum_{b} p(b|x^*,y^*,\lambda) \log p(b|x^*,y^*,\lambda)\right)$ which can be lower-bounded as $H(B|x^*,y^*,\Lambda)\geq \sum_\lambda q(\lambda)f(\Gamma^{(\lambda)})$. 

A bound on $H(B|x^*,y^*,\Lambda)$ can thus be obtained by minimizing $\sum_\lambda q(\lambda)f(\Gamma^{(\lambda)})$ over all ensemble $\{q(\lambda),\Gamma^{(\lambda)},{\Upsilon}^{(\lambda)}_x\}$  of moment and localizing matrices subject to the constraints $\sum_\lambda q(\lambda) \omega_{x,n}^{(\lambda)} = \omega_{x,n}$ and $\sum_\lambda q(\lambda) \alpha_0^{i,(\lambda)} = \alpha_0^{i}$. But any solution  $\{q(\lambda),\Gamma^{(\lambda)},{\Upsilon}^{(\lambda)}_x\}$ defines average moment and localizing matrices $\Gamma = \sum_\lambda q(\lambda) \Gamma^{(\lambda)}$ and ${\Upsilon}_x = \sum_\lambda q(\lambda) {\Upsilon}^{(\lambda)}_x$ that satisfy the constraints coming from (\ref{eq:vn_cc}) \added{and (\ref{eq:identitytrace})} with the same value for the objective function $f(\Gamma) = \sum_\lambda q(\lambda) f(\Gamma^{(\lambda)})$. I.e., by convexity, the SDP relaxaion of (\ref{eq:vn_cc}) \added{and (\ref{eq:identitytrace})} already provides a lower-bound, not only on the Shannon entopy $H(B|x^*,y^*)$ with no-side information, but on the conditional Shannon entropy $H(B|x^*,y^*,\Lambda)$ with classical side information.

\textbf{Explicit SDP relaxation used for our results --} The results and curves presented in the main text have been obtained by considering a moment matrix indexed by operators of the form $S = \{\mathds{1},\{Z_{b,i}\}_{b,i}\}\times\{\{\rho_x\}_x,\{M_{b|y}\}_{b,y},\{\replaced{\sigma_n}{\omega_n}\}_n\}$. Instead of imposing the positivity of the full moment matrix, we imposed positivity of the sub-blocks {$\Gamma_{i}$} indexed by operators {$S_{i} = \{\mathds{1},\{Z_{b,i}\}_b\}\times\{\{\rho_x\}_x,\{M_{b|y}\}_{b,y},\{\omega_n\}_n\}$ involving a list $\{Z_{b,i}\}_b$ indexed by a single $i$. The specific SDP level we solved to obtain the Shannon entropy curves from state discrimination games contains monomials up to second level (namely, $k=2$) in the states, measurements and photon-number operators, i.e.
\begin{align}
S_{i} = \{\mathds{1},\{Z_{b,i}\}_b\}\times\{&\{\rho_x\}_x,\{M_{b|y}\}_{b,y},\{\replaced{\sigma_n}{\omega_n}\}_n,\{\rho_x\rho_{x'}\}_{x,x'},\{\rho_xM_{b|y}\}_{x,b,y},\{\rho_x\sigma_n\}_{x,n},\{M_{b|y}\rho_x\}_{b,y,x}, \\
&\{M_{b|y}M_{b'|y'}\}_{b,y,b',y'},\{M_{b|y}\sigma_{n}\}_{b,y,n},\{\sigma_n\rho_x\}_{n,x},\{\sigma_nM_{b|y}\}_{n,b,y},\{\sigma_n\sigma_{n'}\}_{n,n'}\} \nonumber \ .
\end{align}
The curves bounding the Shannon entropy from the BPSK protocol in the main text are obtained by solving the same SDP level plus two elements from the third level, referred as $k=2+\rho\rho\rho+\sigma M\rho$ in the main text. Specifically,
\begin{align}
S_{i} = \{\mathds{1},\{Z_{b,i}\}_b\}\times\{&\{\rho_x\}_x,\{M_{b|y}\}_{b,y},\{\replaced{\sigma_n}{\omega_n}\}_n,\{\rho_x\rho_{x'}\}_{x,x'},\{\rho_xM_{b|y}\}_{x,b,y},\{\rho_x\sigma_n\}_{x,n},\{M_{b|y}\rho_x\}_{b,y,x}, \\
&\{M_{b|y}M_{b'|y'}\}_{b,y,b',y'},\{M_{b|y}\sigma_{n}\}_{b,y,n},\{\sigma_n\rho_x\}_{n,x},\{\sigma_nM_{b|y}\}_{n,b,y},\{\sigma_n\sigma_{n'}\}_{n,n'}, \nonumber \\
&\{\rho_x\rho_{x'}\rho_{x''}\}_{x,x',x''},\{\sigma_nM_{b|y}\rho_{x}\}_{n,b,y,x}\} \nonumber \ .
\end{align}
In a nutshell, the specific SDP we solve can be written as
\begin{align}
\label{eq:final_SDP}
   H_{x^\ast,y^\ast} = c_m + \sum_{i=1}^{m-1} \underset{\Gamma_i,\Upsilon_{i,x}}{\text{minimize}} & \quad \tau_i \sum_{b} \ \left[2(\Gamma_i)_{Z_{b,i}\rho_{x^\ast},M_{b|y^\ast}} + (1-t_i)(\Gamma_i)_{Z_{b,i}\rho_{x^\ast},Z_{b,i}M_{b|y^\ast}} + t_i (\Gamma_i)_{Z_{b,i}\rho_{x^\ast},Z_{b,i}}\right]  \\
   \text{such that} & \quad \Gamma_{i}\succeq 0 \ , \quad \Upsilon_{x,i}\succeq 0  \nonumber \\
	& \quad (\Gamma_i)_{\rho_x,\mathds{1}} =(\Gamma_i)_{\sigma_n,\mathds{1}} = 1 \ , \quad (\Gamma_i)_{\rho_x,M_{b|y}}=p(b|x,y) \ , \quad (\Gamma_i)_{\rho_x,\sigma_n}\geq 1-\omega_{x,n} \nonumber \\\nonumber
    & \quad + \text{linear constraints on $\Gamma_i$, $\Upsilon_{x,i}$ coming from cyclicity, commutation,}\\
    & \quad \quad\text{ operator equalities and Eq.~\eqref{eq:identitytrace}}\nonumber .
\end{align}
We refer to the the open repository \url{https://github.com/chalswater/Energy_restricted_QRNG} for our implementation which includes the detailed list of constraints.

\section{More fine-grained binnings for the BPSK protocol} \label{App:BPSK}

Here, we give supplementary details on the expected quantum correlations for the BPSK protocol, for more fine-grained binnings. The BPSK protocol \cite{VanHimbeeck2017semidevice} has Alice choosing between two coherent states $\ket{\pm \alpha}$,  with  $\alpha>0$ These states have average photon number $\langle N \rangle = \alpha^2$, and non-vacuum component $1-\exp(\alpha^2)$. Bob performs a homodyne measurement of the $X$ quadrature. To analyse the randomness in the outcome distribution, we define a binning of the continuous outcome into several discrete outcomes $b$. Given a binning, the resulting distribution can then straightforwardly be computed using the identity
\begin{equation}
	\langle x | \alpha\rangle= \frac {e^{-\frac{(x-\sqrt{2}\alpha)^2}{2}}}{\pi^{1/4}} ~.
\end{equation}

The analysis of \cite{VanHimbeeck2017semidevice} considers a binary binning of the continuous output of the homodyne measurement of the $X$ quadrature, $b =1$ if $X<0$ and $b=2$ if $X>0$. With this choice, the probability of obtaining $b=1$ when Alice prepares $\ket{\alpha}$ is equal to,
\begin{equation}
	p(1|\alpha) = \frac{1}{2}\qty(\erf \qty( \sqrt{2} \alpha) + 1) ~.
\end{equation}
The remaining probabilities are fixed by normalisation and symmetry, i.e. $p(2|\alpha) = 1- p(1|\alpha), p(1|-\alpha) = p(2|\alpha), p(2|-\alpha) = p(1|\alpha)$.
Let us now consider a finer-grained binning of the continuous output, into four outcomes. Taking into account the symmetry of the setup. A natural choice is to bin according to 
\begin{equation}
    \begin{aligned}
        X \in (-\infty, -x_1) &\rightarrow b = 1, &
        X \in (-x_1, 0) &\rightarrow b = 2, \\
        X \in (0, x_1) &\rightarrow b = 3, &
        X \in (x_1, \infty) &\rightarrow b = 4.
        \end{aligned}
\end{equation}
where $x_1$ is a positive real number. The probabilities of obtaining each outcome can be computed by integrating the Gaussian distribution over the corresponding intervals, and normalizing the result to one. We find
\begin{equation}
\begin{aligned}
    p(1|-\alpha) &= \frac{1}{2}\qty(\erf \qty(\sqrt{2}\alpha - x_1) + 1), &
    p(1|\alpha) &= \frac{1}{2}\qty(1 - \erf \qty(\sqrt{2}\alpha + x_1)), \\
    p(2|-\alpha) &= \frac{1}{2}\qty(\erf \qty(x_1 - \sqrt{2}\alpha) + \erf \qty(\sqrt{2}\alpha)), &
    p(2|\alpha) &= \frac{1}{2}\qty(\erf \qty(x_1 + \sqrt{2}\alpha) - \erf \qty(\sqrt{2}\alpha)).
    \end{aligned} \end{equation} 
and by symmetry $p(3|\alpha) = p(2|-\alpha)$, $p(3|-\alpha) = p(2|\alpha)$, $p(4|\alpha) = p(1|-\alpha)$ and $p(4|-\alpha) = p(1|\alpha)$. Finally, we choose $x_1$ such that \begin{equation} p(1|-\alpha) = p(2|-\alpha)~. \end{equation} This leads us to take $
	x_1 = \sqrt{2}\alpha-\erf^{-1}\qty(\frac{1}{2} \qty(\erf \qty(\sqrt{2}\alpha)-1)) ~$.

Similarly, in the case of eight outcomes, we bin according to the intervals $(-3x_1,-2x_1,x_1,0,x_1,2x_1,3x_1)$, and compute the corresponding probabilities in the same way. We computed the min-entropy for the BPSK protocol based on several alternative binnings and found that the above choice of $x_1$ leads to the highest randomness rate.

\end{document}